\documentclass[]{eas}
\usepackage{graphicx}
\begin{document}
\runningtitle{The mass distribution in Spiral galaxies}

\title{The mass distribution in Spiral galaxies} 
\author{Paolo Salucci}\address{SISSA, Via Beirut 2, Trieste, salucci@sissa.it}
\author{Christiane Frigerio Martins}\address{UFABC, Rua Catequese 242, 09090-400 Santo Andr\'{e}-S\~{a}o Paulo, Brazil, cfrigerio@ufabc.edu.br}

\begin{abstract}
 In the past years a wealth of observations has unraveled the structural properties of the Dark and Luminous mass distribution in spirals. These have pointed out to an  intriguing scenario not easily explained by present theories of galaxy formation.
The investigation of individual and  coadded objects has shown that the spiral  rotation curves follow, from their centers out to  their virial radii, a Universal profile (URC) that  arises from  the tuned  combination of  a stellar disk and of a dark halo. The importance of the latter component  decreases with galaxy mass. Individual objects, on the other hand, have  clearly revealed  that the  dark halos encompassing the luminous discs  have a constant  density core. This resulting observational  scenario  poses important challenges to  presently favored theoretical  $\Lambda$CDM Cosmology.   
\end{abstract}
\maketitle
\section{Introduction}
 
The presence of large amounts of unseen matter in and around  spiral galaxies, distributed  differently from   stars and gas,  is well established primarily  from optical and 21 cm RCs  which do not show the expected Keplerian fall-off at large radii but remain increasing, flat or gently decreasing over their entire observed range (Bosma 1981, Rubin 1980). The invisible mass component becomes progressively more abundant at outer radii   for the less luminous galaxies (Persic and Salucci 1988, Broeils 1992).
The  distribution of matter in disk systems has rapidly become a benchmark for the theory of galaxy formation.
In particular, the apparent  universality and the dark-luminous coupling  of the  mass distribution is thought to  bear the imprint of the Nature of the  DM  and of the process of galaxy formation (Donato \emph{et al.} 2009, Gentile \emph{et al.} 2009).
It is well known that numerical simulations performed in the   $\Lambda$CDM scenario predict  well-defined  virialized halos leading to specific  density  and velocity profiles (NFW Navarro, Frenk and White 1996). 

In Spirals the   circular velocity $V(r) $ is the equilibrium velocity implied by   their mass distribution. The gravitational potentials of   a spherical stellar bulge, a DM  halo, a stellar disk, a gaseous disc  
$\phi_{tot}=\phi_b+ \phi_{DM}+\phi_*+\phi_{HI}$
lead to:
\begin{equation}
V^2_{tot}(r)=r\frac{d}{dr}\phi_{tot}=V^2_b + V^2_{DM}+V^2_*+V^2_{HI}. \label{eq:3_vel_tot}
\end{equation}
with  the Poisson equation  relating the surface (spatial)  densities to the corresponding  gravitational potentials. 
 
$\Sigma_*(r)$, the surface stellar  density  
is assumed proportional (by the  mass-to-light ratio) to the luminosity 
surface density  (Freeman 1970):
\begin{equation}
\Sigma_{*}(r)=\frac{M_{D}}{2 \pi R_{D}^{2}}\: e^{-r/R_{D}},\label{eq:3_sigma_stars}
\end{equation}
where $M_D$ is the disk mass and $R_D$ is the scale length. $\Sigma_{HI}(r)$  is   directly obtained by observations.
Equation (\ref{eq:3_sigma_stars}) leads to  
$V_{*}^{2}(r)=\frac{G M_{D}}{2R_{D}} x^{2}B\left(\frac{x}{2}\right)$,
where $x\equiv r/R_{D}$, $G$ is the gravitational constant   $B=I_{0}K_{0}-I_{1}K_{1}$,  a combination of Bessel functions.

The  assumption  that  the  rotation curve  of a  spiral  leads to  a fair  measure of its gravitational potential is well justified. In fact: i) in their  very inner regions  the light  well  traces the gravitating  mass (Ratnam and Salucci 2000) and ii) there exists,  at any  galactocentric radii measured in terms of disk lenght-scale $R_n \equiv (n/5)R_{opt}$,  a    {\it radial}  Tully-Fisher relation (Yegorova and Salucci 2007) linking  with  very low scatter  the local rotation  velocity  $V_n \equiv V_{rot}(R_n)$ with the  total galaxy luminosity (see Fig. 1).
\begin{equation}
M_{band} = a_n \log V_n + b_n , \label{eq:RTF}
\end{equation} 
($a_n$, $b_n$ are the slope and zero-point of the relations)
implying that the RC velocity  is a fair measure of the underlying gravitational potential.

\begin{figure} [t!]
\centering
\vskip -0.6cm
\includegraphics[width=8.8cm]{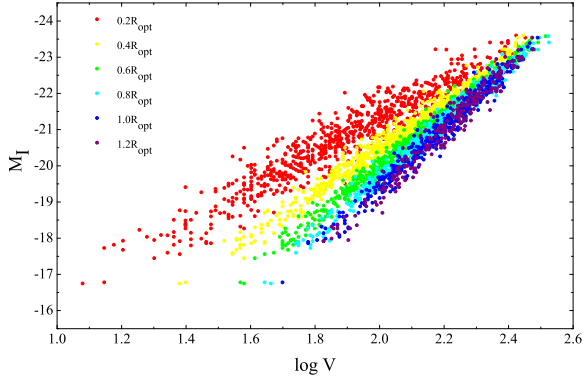}
\vskip -0.7cm
\caption{The Radial TF. The relations at different radii are indicated with different colours.} 
\label{fig:RTF}
\end{figure}

Contrary to what  is  often claimed, the  observational evidence  indicates  that, generally, RCs are  not asymptotically flat \footnote{The interested 
reader may find additional material on the issue by visiting http://www.facebook.com/group.php?gid=310260450630.} 
(see Salucci \emph{et al.} 2007) neither flat inside $R_{opt}$ (Persic and Salucci 1991).
\begin{figure} [t!] 
\centering
\vskip -0.6cm
\hskip -6.9cm
\includegraphics[width=6cm]{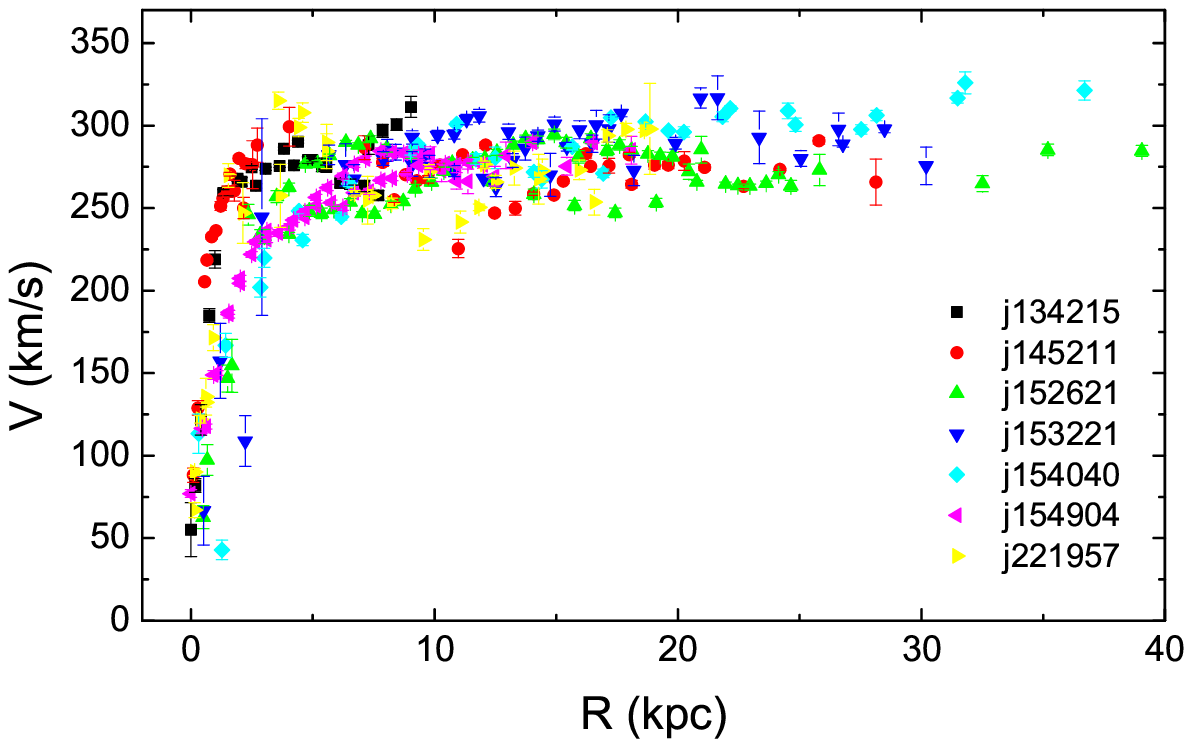}
\vskip -4.2cm
\hskip 6.2cm
\includegraphics[width=3.4cm,angle =-90]{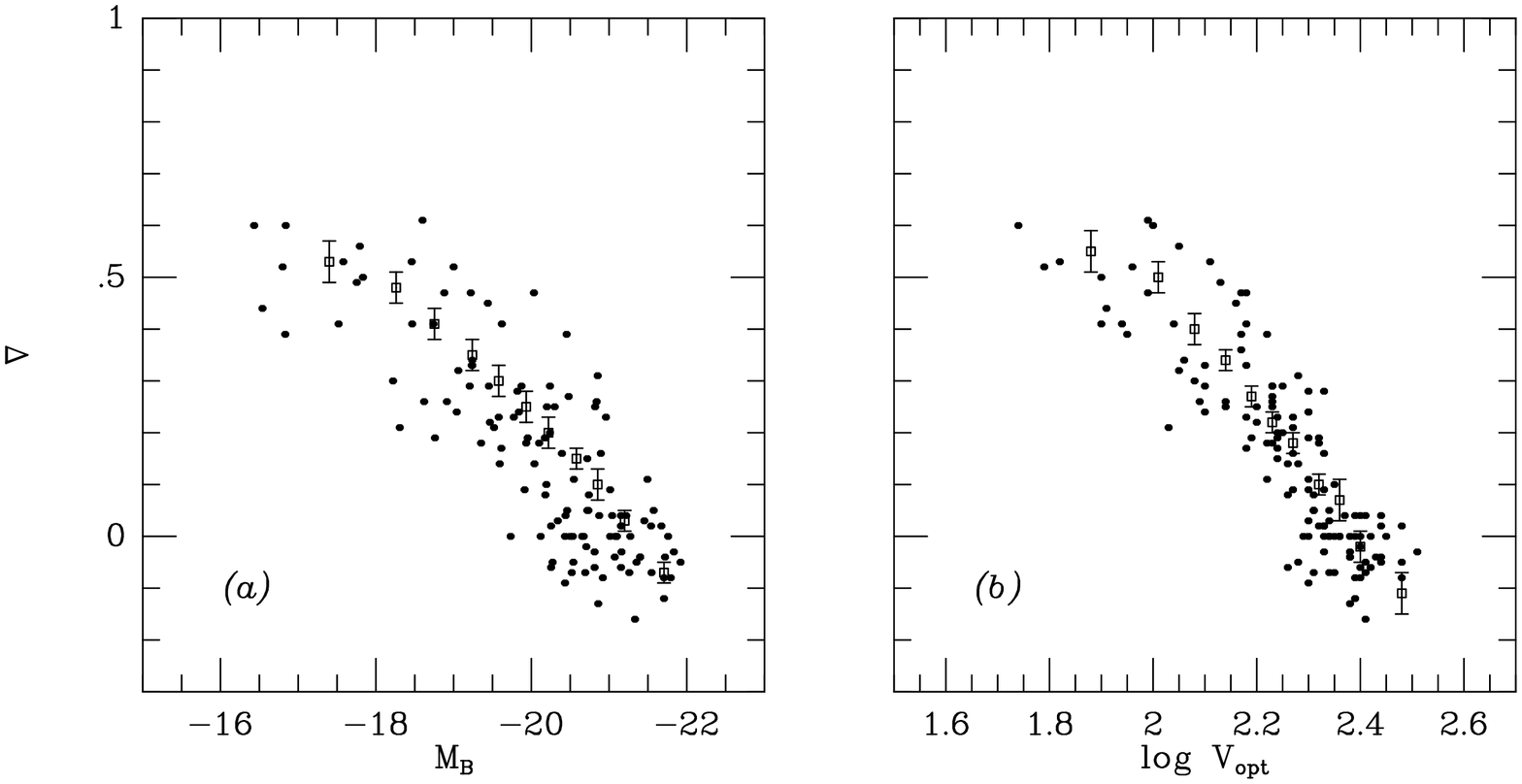}
\vskip 0.2cm
\caption{Left:   RCs of  spirals with $M_B \sim -21.5$.
Right: RC log slope  as a function of $M_B$ and  $ V_{opt}$.} 
\label{fig:non_flat}
\vskip -0.3cm
\end{figure}
In Fig. 2 (right) we plot, for a  large  sample of galaxies, the  logarithmic  {\it slope}  of the circular velocity   at $R_{opt} $ as a function  of $V_{opt}$ and $M_B$. We find: 
 $ -0.2 \leq \nabla \leq 1$: the  RC slopes  take almost  all  the values allowed by Newtonian gravity,   from -0.5 (Keplerian regime) to  1 (solid body regime) and, furthermore, they  strongly correlate  with   galaxy  luminosity and $V_{opt}$  (PSS). 

\section {The Universal Rotation Curve}       

The study  of the systematics of    spiral kinematics,  pioneered by   Persic and Salucci 1991, further developed by PSS and by Salucci \emph{et al.} 2007   has evidenced that the   RCs of spirals  present universal features
(e.g. see Fig. 2 (left)) that, besides, well correlate with global galactic properties. 
This has  led to the construction of the ``Universal Rotation Curve''  $V_{URC}(r; P)$ (see PSS), i.e.  an empirical  function of   galactocentric  radius $r$, that,  tuned by a parameter related to a global galaxy property (e.g. its virial mass),
 is able to reproduce the RC of any object.   A proof of this comes from the 11 coadded (synthetic)  curves $V_{coadd}(r/R_{opt}, M_I)$  obtained  by  binning  616 RCs of late type spirals extended  out to $R_{opt} \equiv 3 R_D $. The resulting curves  are  the   synthetic  RCs of spirals  at  11 fixed  luminosities,  spanning the $I$-band range: $-16.3 < {\rm M}_I< -23.4$. They      result regular and  smooth  with a very   small intrinsic variance and with  a remarkable luminosity dependence.  Additional  kinematical  data,  including    very extended  individual RCs  and     virial velocities $V_{vir}\equiv (G M_{vir}/R_{vir})^{1/2}$ obtained by  Shankar \emph{et al.} 2006,   also contribute  to support the  URC paradigm and to  determine the actual velocity function (Salucci \emph{et al.} 2007).  Then, $V_{URC}$  takes the role of the  observational counterpart of the   velocity profile that emerges  out of numerical cosmological simulations. It  is obtained  by mass  modelling  spirals with    a Freeman disk  and a  DM halo  with  an (empirical) Burkert profile:  

\begin{figure}[t!]
\centering
\vskip -0.5truecm
\includegraphics[width=6.2truecm]{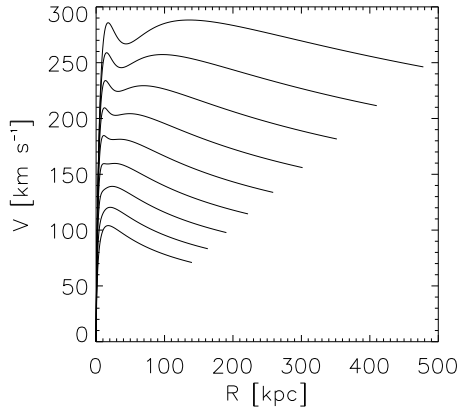}
\vskip -0.5truecm
\caption{The URC.
Each curve corresponds to a galaxy with mass  $M_{vir}=10^{11}  10^{n/5} \rm M_\odot $, with $n = 1 \ldots 9$ from the lowest to the highest curve. } 
\label{fig:io_urc1}
\end{figure}
 
\begin{equation}
\rho (r)={\rho_0\, r_0^3 \over (r+r_0)\,(r^2+r_0^2)}. \label{eq:burkert}
\end{equation}
 $r_0$ is the core radius and $\rho_0$ the central density. We have  $V^2_{URC} = V^2_{URCD} + V^2_{URCH}$, where  $V_{URCH}(r)$  is given  by Eq. 2 of  Salucci and Burkert 2000.

\section {The mass model of Spirals}
  
The  three parameters of the URC   $\rho_0$,  $r_0$,  $M_D$  are determined  by $\chi^2$ fitting    the above-specified  kinematical  data with  the URC mass  model.  The fits  are excellent  (PSS), in addition,   this very  same  $V_{URC}$  fits  equally well  also  individual  RCs. {\it Independent}  support for the URC  paradigm is given in  Salucci \emph{et al.} 2007. As general result, a number   of  scaling laws  among   the structural mass parameters $\rho_0$, $M_D$, $M_{vir}$, $r_0$   emerges  (see Fig. 4, and  Fig. 11 of  PSS). 

\begin{figure}[t!]
\centering
\vskip -0.7cm
\includegraphics[width=9.5truecm]{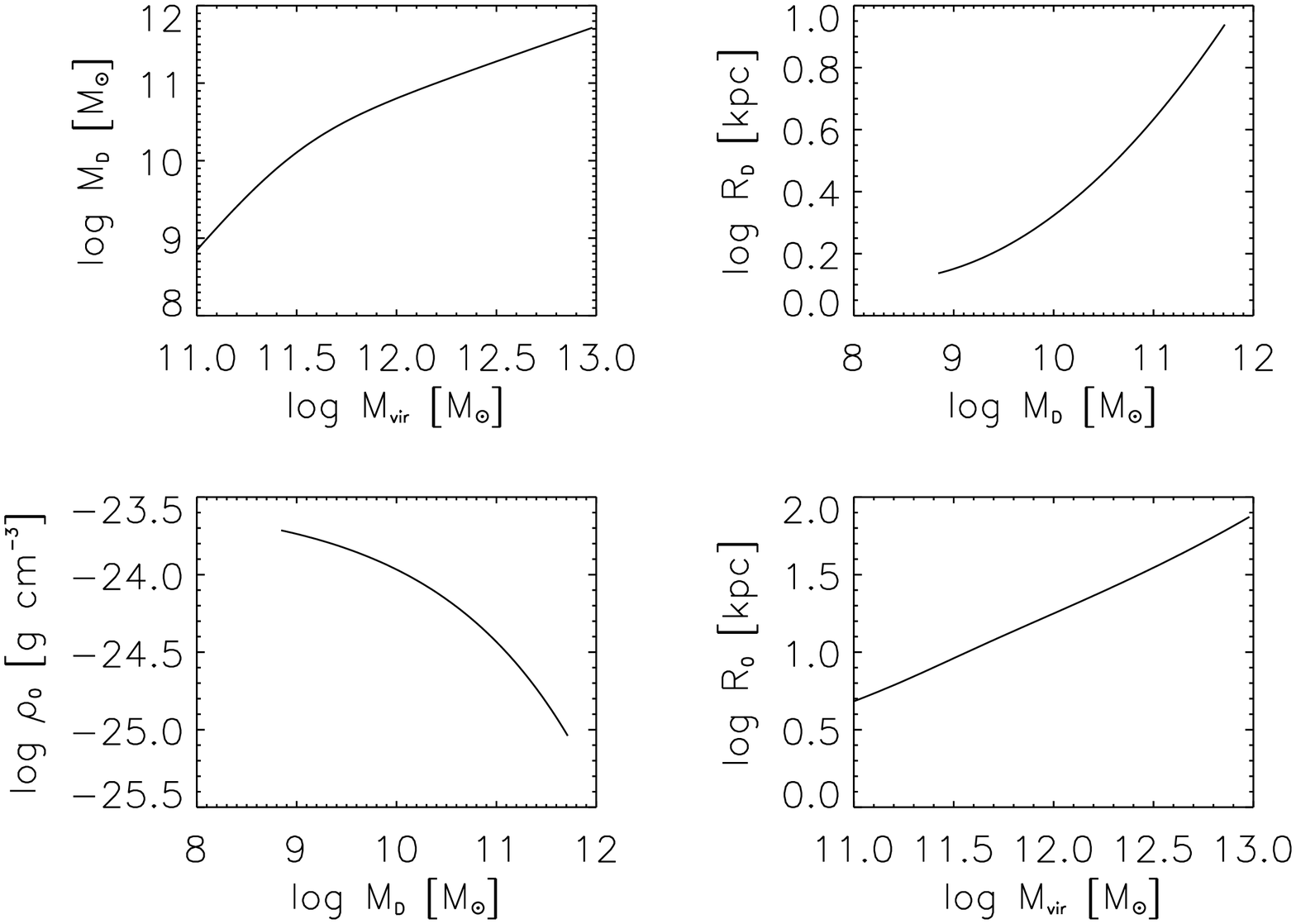}
\vskip -0.7truecm 
\caption{Scaling relations between the structural parameters of spirals.}
\label{fig:scaling_relations}
\end{figure}

These scaling laws indicate (Salucci \emph{et al.} 2007)  that spirals have  an Inner Baryon Dominance region where the stellar disk  dominates  the total gravitational potential,  with the  DM halo  dominating farther out.  At any radii, objects  with lower luminosities have a larger  dark-to-stellar mass ratio. The baryonic fraction in a spiral  is always much smaller than the cosmological value $\Omega_b/\Omega_{matter} \simeq 1/6  $, and it ranges between $7\times 10^{-3}$   to   $5\times 10^{-2}$,  suggesting that processes such as  SN  explosions  must have   removed (or prevented the  transformation in stars of)  a very large fraction of the original hydrogen.
Smaller spirals are denser, with their  central density spanning 2 order of magnitudes over the mass sequence of spirals.
The stellar mass-to-light ratio   (in the B band)   lies  between 0.5 and 4 and increase with galaxy  luminosity  as $L_B^{0.2}$;  in  agreement  with the values   obtained by fitting the spirals SED with spectro-photometric models (Salucci, Yegorova and Drory 2008 and references therein). Dark and  luminous matter in spirals   emerge as  extremely well linked together.

\section{The core-cusp controversy}

The $\Lambda CDM$  scenario provides a successful picture of the cosmological  structure  formation (e.g.  Ostriker 1993).  Large N-body  numerical simulations performed in  this scenario have provided us with the structural properties of the  DM  virialized halos  sourrounding disks today. The halo  spatial density  is found  universal and   well reproduced by one-parameter  radial profile, known as  the Navarro, Frenk \& White (1996) profile $\rho_{DM}$
\begin{equation}
\rho_{NFW}(r) = \frac{\rho_s}{(r/r_s)\left(1+r/r_s\right)^2},
\label{eq:nfw}
\end{equation}
where $r_s$ is a characteristic inner radius, and $\rho_s$ the corresponding density. We remind that  for the virial radius  $R_{vir}$ and  halo mass $M_{vir}$ and  the mean  universal  density $\rho_u$ we have: $
M_{vir}  \simeq   100  \rho_u R_{vir}^3.$ Simulations   show that $r_s$ and $\rho_s$  are related, within a reasonable scatter:    $ R_{vir} / r_s \simeq 9.7 \left( \frac{M_{vir}}{10^{12}M_{\odot}} \right)^{-0.13}$.  
 
It has been  known, since  they were discovered, that cuspy NFW density profiles   disagree with the actual halo  profiles  around   spirals and LSB (Moore 1994, Kravtsov 1998, McGaugh and de Blok 1998, Marchesini \emph{et al.} 2002).
However,   early studies were possibly  affected by   theoretical and observational  systematics.  Later,  e.g. in  Salucci \emph{et al.} 2003 and  Gentile \emph{et al.} 2004  it was realized that the   resolution  of this  riddle  concerning   the very nature of DM relied in   a proper  investigation of (a number of) suitable test-cases.  That is,  a careful analysis  of  2D,   high quality,  extended,     regular, free from deviations from axial symmetry and   homogeneous RCs proved to be reliable up to their second derivative.     
In  all examined  cases,  NFW halo predictions and observations  have been  found 
in plain disagreement on several  aspects: the disk + NFW halo mass model 
{\it i)} fits the RC  poorly and   {\it ii)} implies  an implausibly  low stellar mass-to-light ratio  and in some case {\it iii)} an unphysical  high halo mass (e.g. Spekkens 2005, Gentile \emph{et al.} 2004, 2005, 2007,  Simon 2005, Spano \emph{et al.} 2007, de Blok 2008 and  de Naray \emph{et al.} 2008).

\begin{figure}[t!] 
\centering
\vskip -1.cm
\hskip -6.5cm
\includegraphics[width=7.5cm]{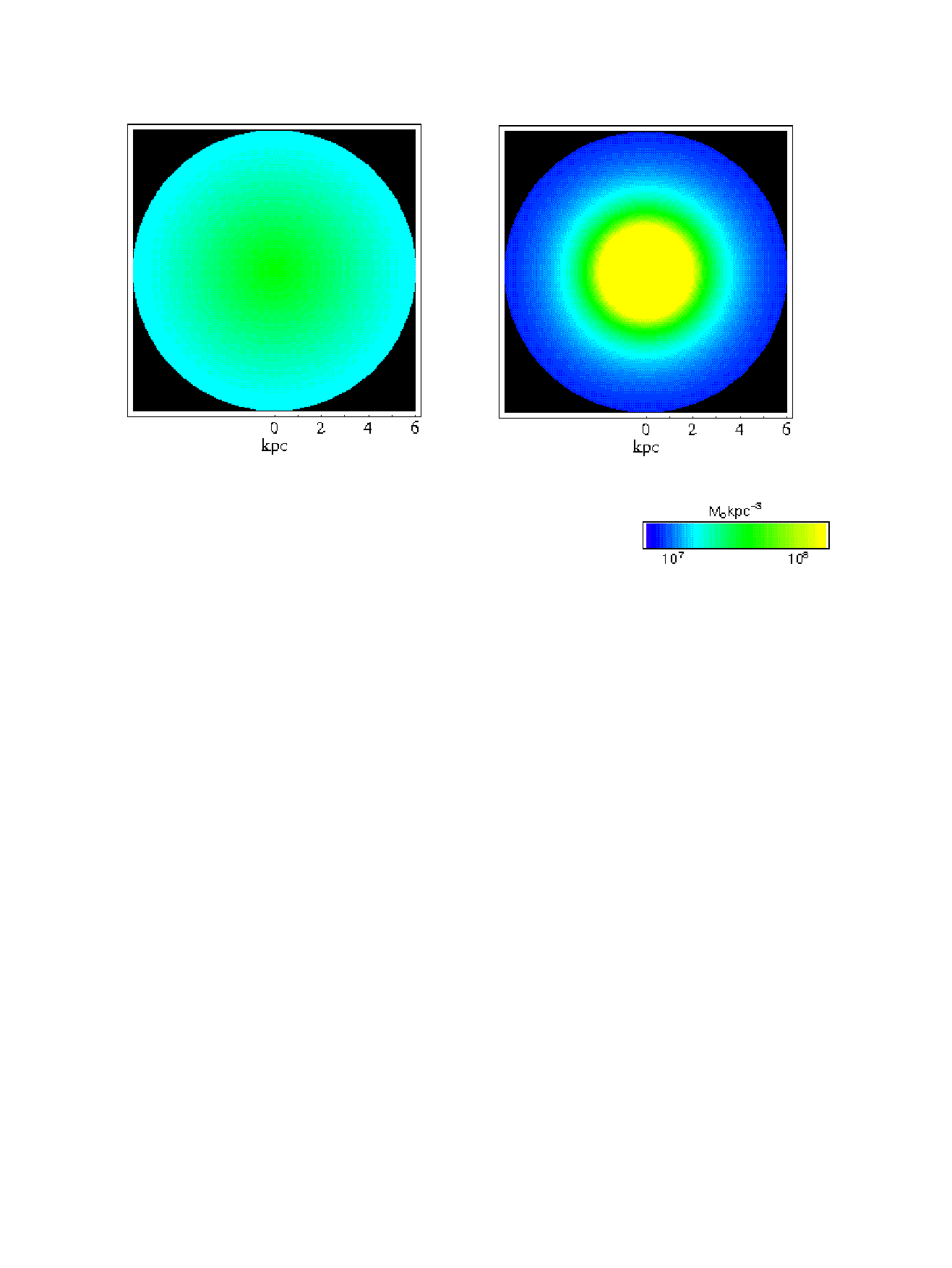}
\vskip -9.7cm
\hskip 7cm
\includegraphics[width=5.2cm]{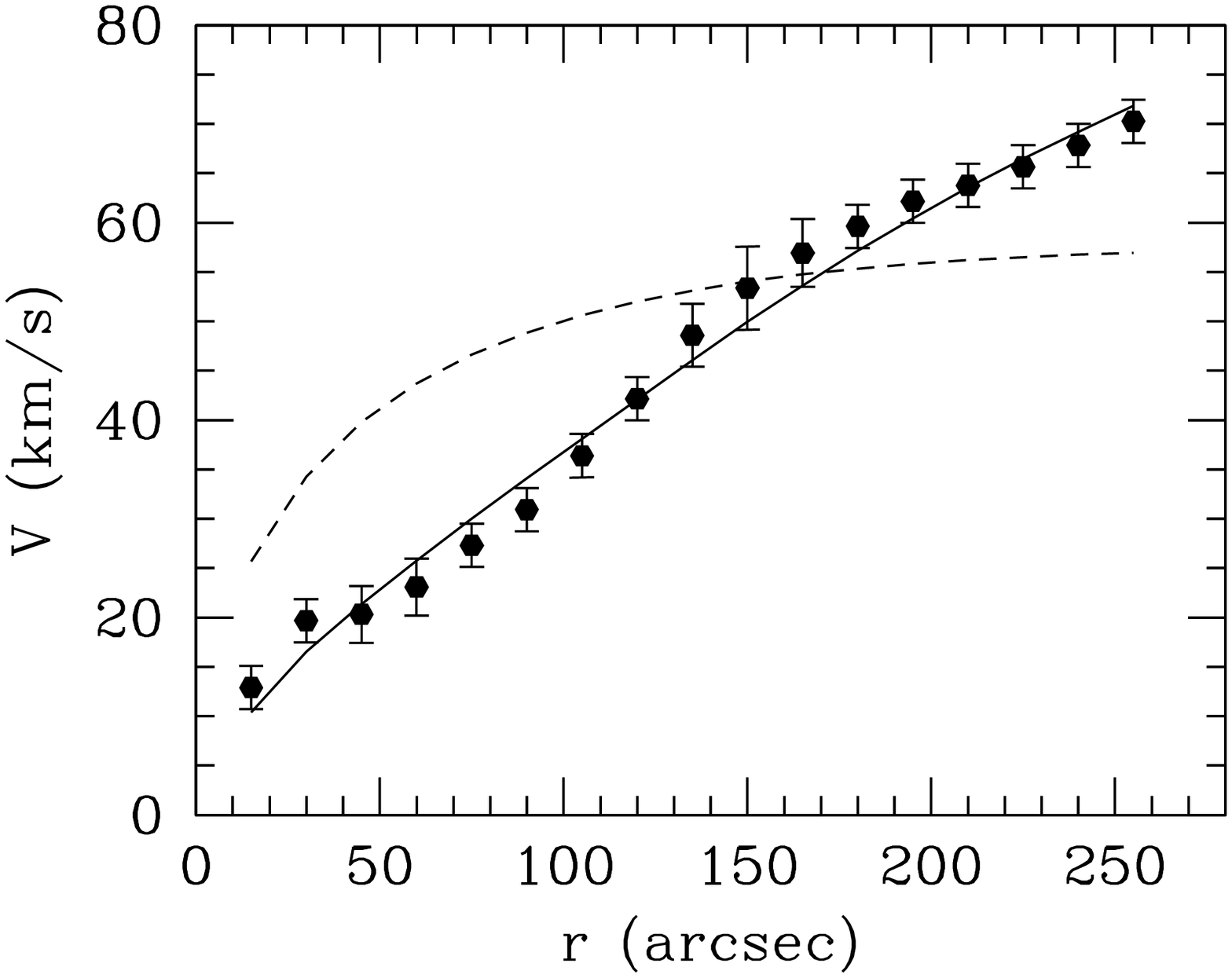}
\vskip -0cm
\caption{Left: Dark halo density in ESO 116-G12: observations (left)  vs. CDM predictions.
Right: RC best-fits of DDO 47: Burkert halo + stellar disk (solid line),  NFW  halo + stellar disk (dashed line)}
\label{fig:cuspVScore}
\end{figure} 

It is worth  illustrating one  example of this disagreement:  the nearby spiral dwarf galaxy DDO 47 (Gentile \emph{et al.} 2005).
The RC mass modeling, shown in Fig. 5, finds that the dark halo has a core radius of about 7 kpc and a central density $\rho_0 = 1.4 \times 10^{-24}$ g cm$^{-3}$:  the underlying DM  density profile is  {\it much} shallower  than that predicted by a NFW profile, totally unable to fit the RC. Presently,  there are  about 50 spirals  whose RCs cannot be reproduced by a NFW halo + a stellar disk. Furthermore, this evidence is accompanied by investigations  that have ruled out that it may  arise from (neglected)  systematical  effects (Gentile \emph{et al.}  2005,   Trachternach  \emph{et al.} 2008, Oh  \emph{et al.}  2008).

 The mass modelling  is not the only far-reaching  theory-data discrepancy present in disk systems. A  complementary evidence comes from    Salucci 2001 in which it was   derived ,  in a model independent way, the logarithmic gradient of the {\it halo}  circular velocity $\nabla_h(r)\equiv \frac{d \log V_h(r)}{d \log r}$ at $R_{opt}$ for  140 spirals of different luminosity (see Fig. 6); their values  $\sim 1$  turned out to  independent of  galaxy magnitude  and inconsistent with  NFW halos predictions. A  similar result  was obtained by  de Blok and Bosma 2002  (see Fig. 6) for  a large sample of  LSB.  
 
The  accurate mass modeling of the external regions of 37 spirals with high quality RCs (Donato \emph{et al.} 2007) led  to the discovery of a new observational feature which is in contrast with the  NFW  predictions.
 The DM halos around spirals that have, in the inner regions,  densities  up to  one order of magnitude {\it lower}  than the $\Lambda$CDM predictions. At about $2-3 R_{opt} $ they  have, instead,  densities   {\it higher}  by a factor 2-4 the  corresponding NFW   predictions  (see Fig. 7). $\Lambda$CDM  halo profiles are  steeper,   at 5 kpc scale, and  shallower, at 50 kpc scale, than the  actual DM halos profiles.   
 
\begin{figure}[t!] 
\vskip -0.3cm
\hskip 0.5cm
\includegraphics[width=9.4cm]{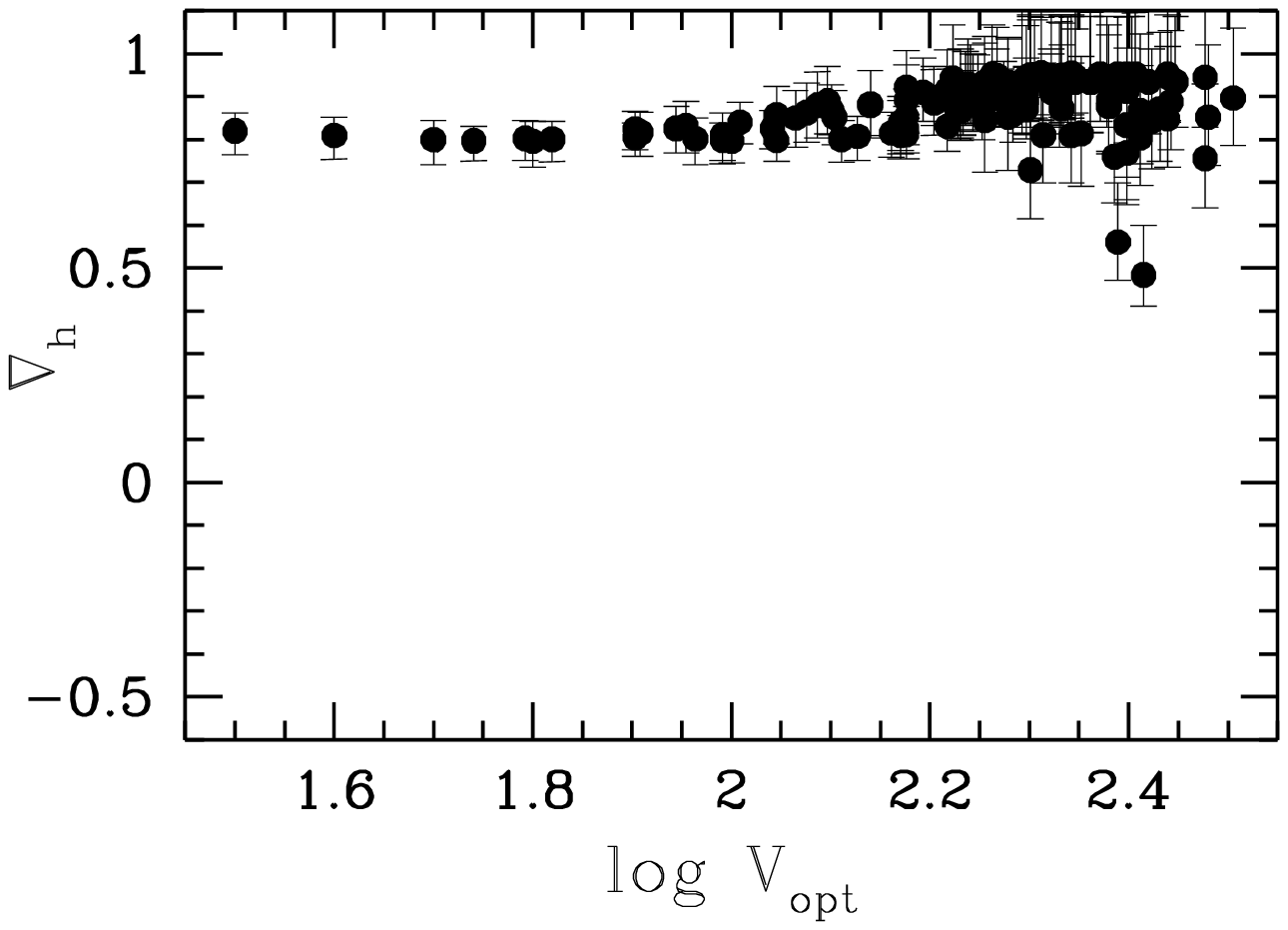}
\vskip -9.4cm
\hskip 7.2cm
\includegraphics[width=5cm]{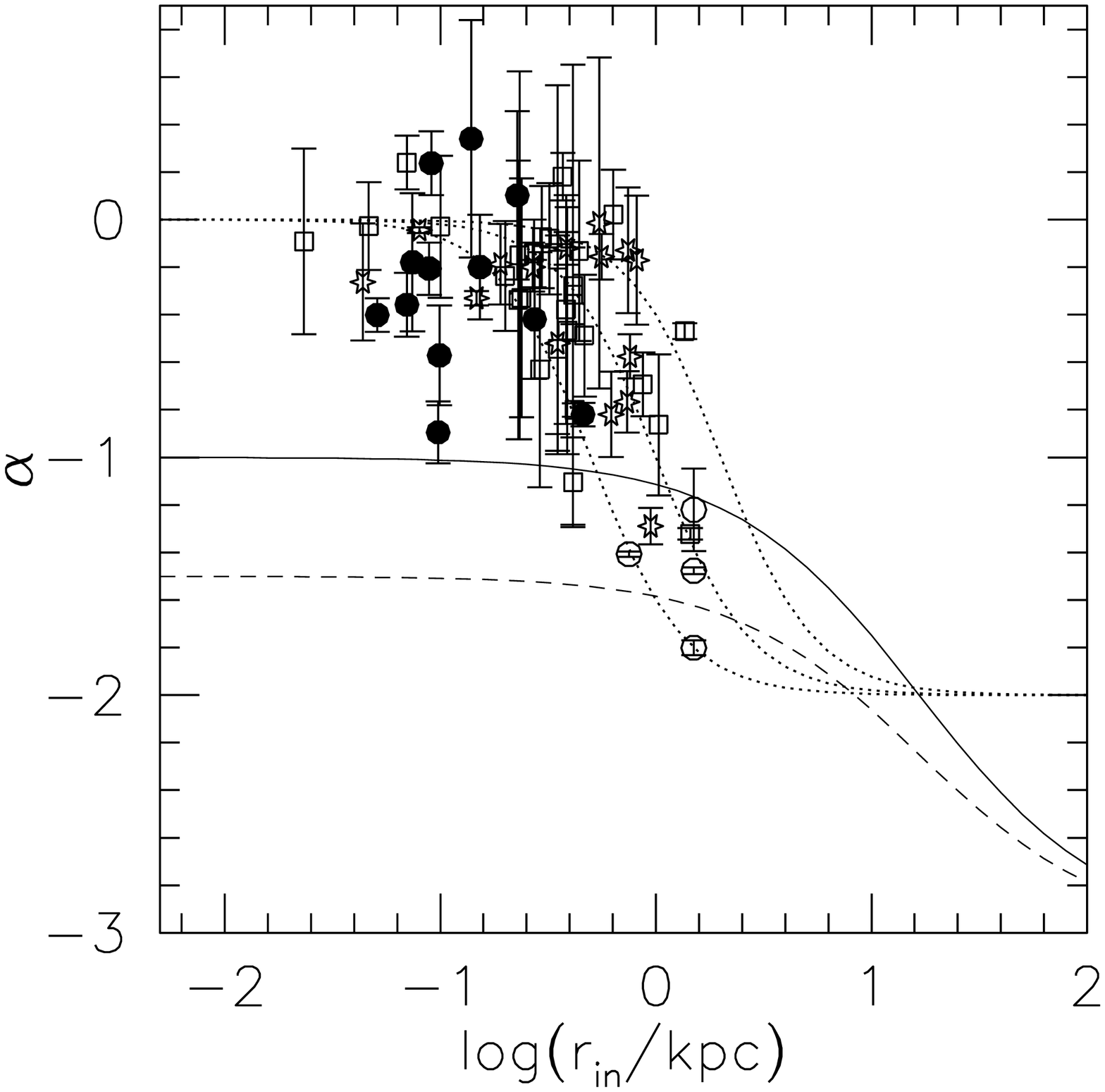}
\vskip -0.0cm
\caption{Left: DM  halo velocity slope  as a function of $V_{opt}$, remind that:    $\nabla_{NFW} \leq 0.3$.
Right: Inner slopes of  LSB halo density  profiles {\it vs}  radii of the innermost points. In comparison,   the slopes of  pseudo-isothermal halo models (dotted lines) with core radii of 0.5, 1, 2  kpc; full and dashed  lines represent the NFW and Moore profiles.  }
\vskip -0.2cm
\label{fig:nablaH}
\end{figure} 

\begin{figure}[b!]
\centering
\vskip -0.6cm
\includegraphics[width=4.8cm]{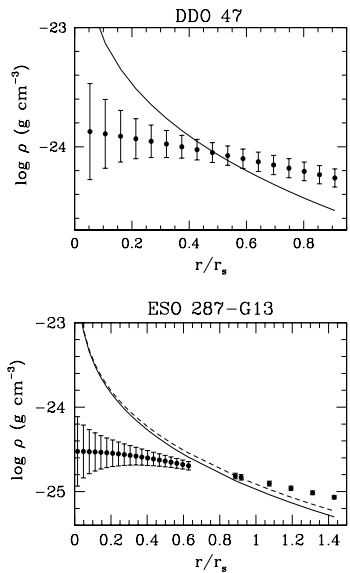}
\vskip -0.2cm
\caption{DM density profiles for DDO 47 and ESO 287-G13 ({\it circles}),  
{\it Solid/dashed }lines: best fits for a  NFW density profile. }
\label{fig:rho_47_287}
\end{figure}

\section{Final remarks: the intriguing evidence}
These discrepancies  have triggered   many alternatives to the strict $\Lambda$CDM-NFW paradigm, some  of them related  to  the process of galaxy      formation and   some to  the very  nature of the dark  particle. Furthermore, also  new   fundamental physics has been invoked.
From an empirical point of view, the distribution of luminous and dark matter  in galaxies shows amazing properties and a  remarkable systematics that  is bound to  play a  decisive role  in shaping    how    $\Lambda$CDM-based theories of galaxy formation must be modified  in order  to  meet with  the challenges that  observations   pose.

 \end{document}